# Ultra-pure RF tone from a micro-ring resonator based optical frequency comb source

Alessia Pasquazi,[1] Marco Peccianti,[2] Brent E. Little,[3] Sai T. Chu,[4] David J. Moss,[5] and Roberto Morandotti[1]

[1]*INRS - Énergie, Matériaux et Télécommunications, 1650 Blvd Lionel Boulet, Varennes (Québec), J3X1S2, Canada*
[2]*Institute for Complex Systems, CNR, Via dei Taurini 19, 00185 Roma (Italy),*
[3]*Infinera Ltd, 169 Java Drive, Sunnyvale, California 94089, USA*
[4]*City University of Hong Kong, Department of Physics and Material Science, Tat Chee Avenue, Hong Kong, China*
[5]*Centre for Ultra-high Bandwidth Devices for Optical Systems (CUDOS) and the Institute of Photonics and Optical Science (IPOS), School of Physics, University of Sydney, Sydney, NSW 2006, Australia*
*peccianti@gmail.com

**Abstract:** We demonstrate a novel mode locked ultrafast laser, based on an integrated high-Q micr-oring resonator. Our scheme exhibits stable operation of two slightly shifted spectral optical comb replicas. It generates a highly monochromatic radiofrequency modulation of 60MHz on a 200GHz output pulse train, with a linewidth < 10kHz.

**1. Introduction**

The demand for high repetition rate, ultrashort pulsed lasers [1,2] is being driven by many applications from next generation time-domain multiplexed optical networks [3,4] to multiple wavelength sources for optical interconnects [5,6], arbitrary all optical waveform generation [7], on-chip frequency combs [8,9], and many others. Passively modelocked fiber lasers, free of the bandwidth bottleneck of electronics, represent an attractive solution to the increasing demand for high repetition rate sources. High repetition rates have been achieved in very short laser cavities with large frequency mode spacings (i.e., large Free Spectral Range (FSR)) [1,10-11], by simply reducing the pulse round-trip time. Conversely, fiber based designs usually rely on approaches where multiple pulses are produced for each round trip [12-14]. As proposed by Yoshida et al. in 1997, the repetition rate of a passive mode-locked fiber laser can be controlled by introducing a Fabry Pérot (FP) filter in the main cavity. The FP suppresses all but a few modes that are periodically spaced with the (inverse) FSR of the FP, leading to a train of pulses with a controlled, and high, repetition rate. The main cavity modes selected by the filter exchange energy via four wave mixing (FWM) and lock their mutual phase as a traveling pulse emerges. This approach, subsequently interpreted in terms of dissipative Four-Wave-Mixing (FWM) [15,16], and variations of it, has been used to demonstrate high repetition-rate pulse trains [16-19]. However, the stability of dissipative FWM schemes still remains a severe issue [1,18] - the FP finesse is typically such that many main cavity modes can oscillate within a single FP filter bandwidth. Their beating results in severe low frequency noise that produces extremely unstable operation [18]. For this reason, this scheme is only marginally suitable for practical applications.

Recently [20] we reported the first stable mode-locked laser achieved by introducing a novel variation of dissipative FWM that we termed *Filter-Driven Four Wave Mixing* (FD-FWM). The key to the success of this approach was to combine the linear filter with the nonlinear element [5] via an integrated nonlinear monolithic high Q (quality) factor micro ring resonator, similar in spirit to optically pumped multiple wavelength oscillators based on high Q-factor resonators [5,6,8,9,21,22]. The high efficiency of the nonlinear wave mixing in the ring resonator eliminates the need for the long external cavities required in earlier approaches. Our scheme relies on optical nonlinearities only within the micro-resonator. Therefore it allows much shorter main cavity lengths and hence wider main cavity mode frequency spacings. This in turn restricts the number of oscillating main cavity modes within a single nonlinear micro ring resonance to at most very few - potentially even a single mode - and thus removes the inherent source of super mode instability. Using this approach we achieved [20] extremely stable operation at high repetition rates while maintaining very narrow linewidths, thus leading to a very high quality pulsed emission. Moreover, the FD-FWM scheme is also intrinsically capable of producing much narrower linewidths than ultrashort cavity mode-locked lasers oscillating at similar repetition rates because the long main cavity results in a much smaller Schawlow-Towns phase noise limit [23,24]. Yet a further key advantage of this approach is that it is highly robust to external (i.e., thermal) perturbations - a well-known problem in resonator-based optical parametric oscillators (OPOs) [7,9] where temperature variations detune the microresonator from the external pump source. Although thermal locking can address this problem, it is ineffective against slow temperature drifts, and often results in the OPO shutting down [25].

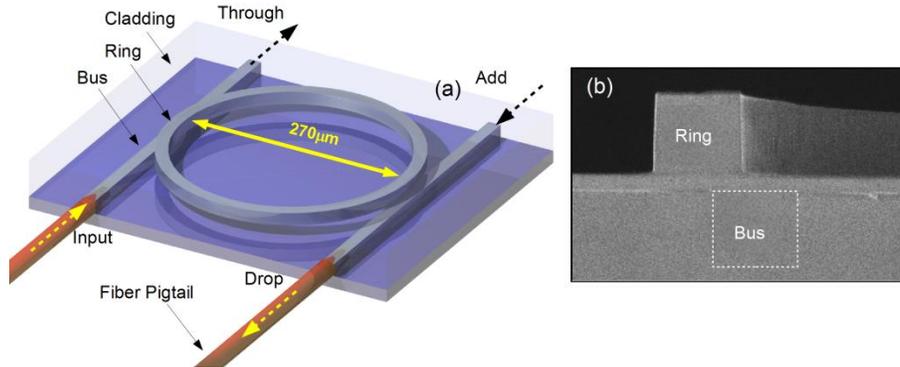

Figure 1. a) Ring resonator microcavity. b) SEM picture of the ring cross-section before depositing the upper cladding of $SiO_2$. The waveguide core is square with transverse dimensions $1.45\mu m \times 1.45\mu m$ and is made of high index (n=1.7) doped silica glass.

Our approach avoids this problem since the oscillating lines are modes of the main cavity that includes the nonlinear resonator and so thermally induced changes in the resonator also change the main cavity length, compensating for the frequency misalignment between the two. Moreover, the increase in overall efficiency obtained with our approach allows oscillation at power levels in the 10mW range, about an order of magnitude lower than OPOs, thus further reducing resonator heating issues.

In this paper, we demonstrate a new stable operating regime for the FD-FWM scheme where *two* main cavity modes within each microresonator linewidth are allowed to oscillate. This novel operating regime leads to the formation of two spectral comb replicas separated by the free-spectral range (FSR) of the external main cavity which in this case is $FSR_{MC}$= 65MHz. The beating of the two combs generates a sinusoidal modulation of the output pulse train at the radio-frequency of the main cavity FSR, and being in the RF regime is readily detectable with standard photodiodes and RF equipment. In addition to providing a high fidelity, ultra-stable RF tone, this approach yields key information about the phase noise and relative modal frequency coherence of this type of laser that is otherwise only accessible using very difficult and indirect optical measurement methods.

## 3. Experiment

The integrated resonator geometry as well as a SEM image of the device cross section is shown in Figure 1. The resonator is a high Q-factor micro-ring (FSR=200 GHz, Q=$1.2\times10^6$) with a 160MHz linewidth [5, 26]. The waveguide core consists of low-loss, high-index (n=1.7) doped silica glass, buried within a $SiO_2$ cladding, and was fiber pigtailed to a standard SMF fiber with a typical coupling loss of 1.5dB/facet. The waveguide cross section is 1.5 µm x 1.45µm while the ring radius is 135µm. The advantages of this platform reside in its negligible linear (< 6dB/meter) and nonlinear losses as well as in a nonlinear parameter as high as $\gamma \sim 220W^{-1}km^{-1}$ [27-29].

Figure 2 shows the laser configuration. The microring resonator was simply embedded in a fiber loop cavity that contained a short erbium doped fiber amplifier, acting as the gain medium. The loop also contained a passband filter with a bandwidth large enough to pass all of the oscillating lines, with the main purpose of controlling the central wavelength $\lambda_0$, as well as an optical isolator, a polarization controller, and an output beam sampler embedded in a delay line of length $L_d$. The total main cavity length was 3m yielding an FSR of 65MHz. The delay line was employed to control the phase of the main cavity modes with respect to the ring modes. By adjusting the delay line of our system we observed three main operation regimes: *unstable*, *stable* and *dual mode*, described in detail in the next section.

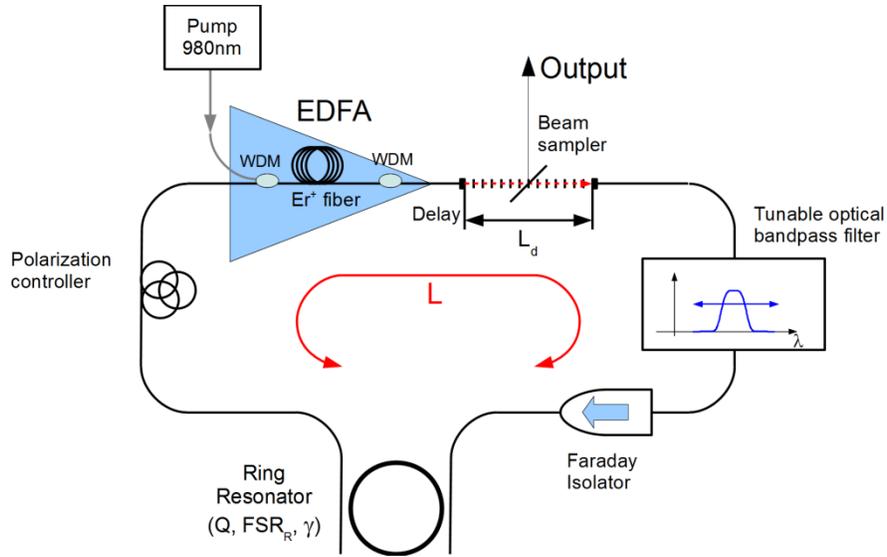

Figure 2. Laser configuration. The free space delay line was used to adjust the spectral position of the main cavity modes relatively to the ring resonator modes.

## 3. Results

*3.1 Unstable Laser Operation.*

Fig. 3(a-d) summarizes the results of the unstable pulsed operation of the laser. This oscillation was obtained by detuning the main cavity delay line and driving the EDFA at the maximum current available in our system, corresponding to approximately 16mW of average optical power at the input of the microring. The optical spectrum recorded with an OSA in Fig. 3(b) shows a comb-like spectrum spaced by the FSR of the microring, in turn indicating

that a fast modulation of the laser output is present, also confirmed by the 200GHz periodic modulation of the signal in Fig. 3(a) - obtained with a second harmonic (SH) non-collinear autocorrelator (black). However, the contrast in the autocorrelation is extremely poor. Indeed, the background in autocorrelation measurements offers a good indication of the quality of the laser pulsed output[7,20]. A non-collinear SH autocorrelation of an isolated pulse is background free. Conversely, the background of the autocorrelation of a coherent train of pulses is dependent on the pulse length vs. repetition rate but can be easily inferred from the OSA spectrum assuming that the lines in Fig.3(b) are fully coherent (all in-phase). The calculated autocorrelation under this assumption is shown in Fig. 3 (a) (dashed black line). The discrepancy between the calculated and measured autocorrelation traces indicates that the mode-locking condition is indeed not guaranteed - random beating between the modes in the laser produces a large average background and results in an unstable operating regime.

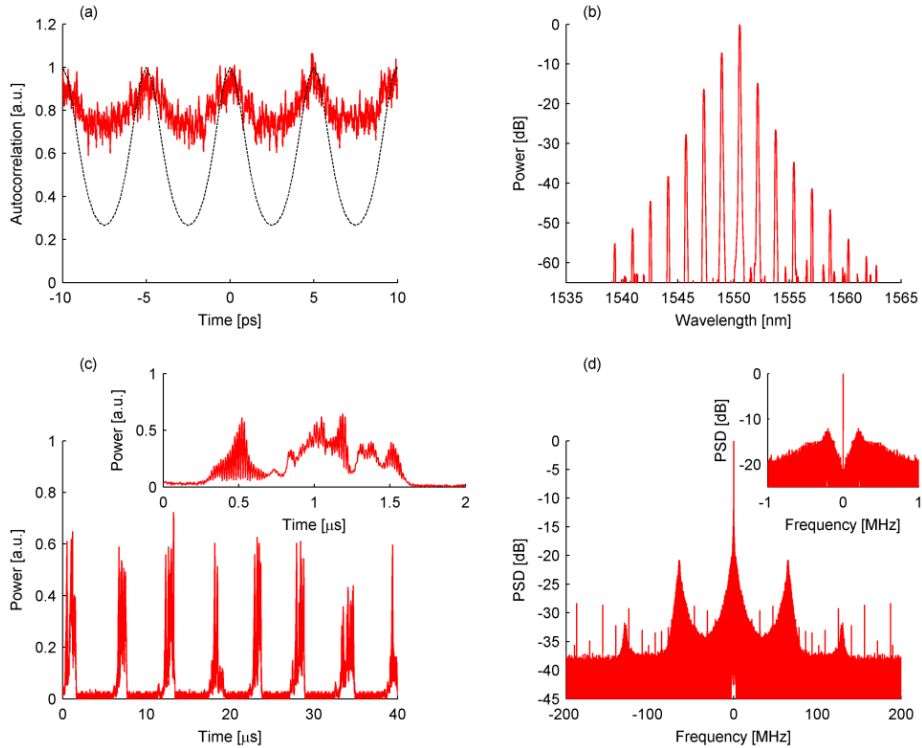

Figure. 3. Optical (a-b) and Radio-Frequency (c-d) characterization of the laser output in the unstable regime: (a) experimental autocorrelation trace (red) and (b) experimental optical spectrum. The autocorrelation trace for a fully coherent transform limited system calculated from the spectrum in (b) is shown in dashed black in (a). RF signal of the laser output in time (c) and spectrum (d). The signal shows an irregular pulsation in the μs scale, as visible in (c) and in the inset of (d), where a zoom of the DC spectral components is plotted. In the inset in (d) a peak of the low-frequency noise components is visible around 150kHz. The RF signal also shows a pulsation due to the beating of the main cavity modes (super mode instability), as visible in the spectral components at 65MHz in (d), and in the fast modulation in the inset in (c).

To have a better insight into the low frequency behavior of the laser, we recorded the envelope signal by using a 150MHz bandwidth photo-detector. The traces are shown in Fig. (c-d) in time and frequency, respectively. We observed a temporal behavior characterized by two temporal scales: (i) a *fast scale* due to the main cavity mode beating, as clearly visible in the spectrum, showing two lateral wings around the FSR of the main cavity (65MHz) and (ii) a *slow scale*, imposing a modulation having period of the order of 6μs, that indicates that the cavity is working in an unstable Q-switching regime, due to the slow recovery time of the EDFA gain, a common occurrence in fiber lasers [30].

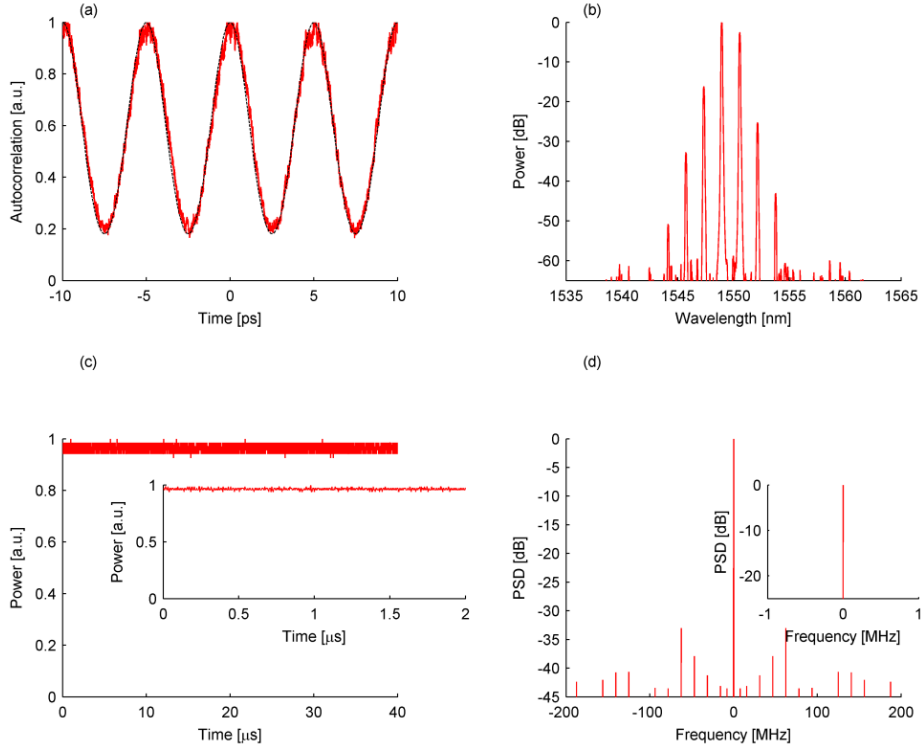

Figure. 4. Optical (a-b) and Radio-Frequency (c-d) characterization of the laser output in the stable pulsed regime: (a) experimental autocorrelation trace (red) and (b) experimental optical spectrum. The autocorrelation trace for a fully coherent transform limited system calculated from the spectrum in (b) is shown in dashed black in (a): the near perfect agreement with the experimental trace indicates that the lines are indeed mode-locked. RF signal of the laser output in time (c) and spectrum (d). The insets in (c) and (d) are zoom-ins of the temporal trace and of the DC component of the spectrum, respectively.

*3.2 Stable Pulsed Operation.*

By adjusting the delay line it was possible to completely eliminate any main cavity low-frequency beating, allowing the laser to easily be stabilized and yielding the very clean results presented in Fig. 4. Several stable oscillation conditions were found via tuning the delay by over 2 cm. Note that the results in Fig. 4 were obtained with the same gain as in the unstable case, leading to an average optical power of approximately 16mW at the microring input in both cases. We note that the optical bandwidth associated with unstable operation (Fig. 3(b)) was generally wider than the stable case (Fig. 4 (b)), because the instability resulted in amplitude modulations of the optical pulse train in the main cavity due to unstable Q-switching, thus increasing the statistical peak power and enhancing the nonlinear interactions. On the other hand, the autocorrelation trace of the stable operation regime corresponds to that predicted for a fully coherent transform limited system (shown in dashed black in Fig. 4 (a)), thus nearly perfectly matching the measured trace (presented in red in Fig.4 (a)). The RF trace (c-d) is a stable, continuous signal, indicating that in this case only a single main cavity mode is oscillating at the center of any of the ring resonances. A complete characterization of the laser in the stable operating regime and as a function of the optical gain is discussed in detail in [20].

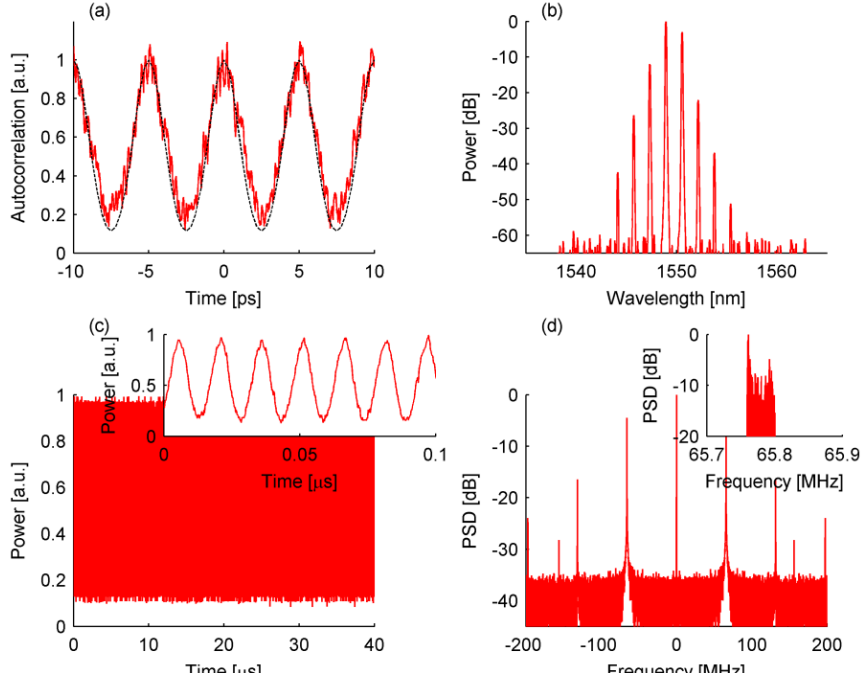

Figure. 5. Optical (a-b) and Radio-Frequency (c-d) characterization of the laser output in the stable pulsed regime under dual comb operation: (a) experimental autocorrelation trace (red) and (b) experimental optical spectrum. The autocorrelation trace for a fully coherent transform limited system calculated from the spectrum in (b) is shown in dashed black in (a): the near perfect agreement with the experimental trace indicates that the lines are in a mode-locked regime. RF signal of the laser output in time (c) and spectrum (d). The insets in (c) are a zoom-in of the temporal trace, showing a monochromatic beating at 65.8 MHz. The inset in (d) shows the AC component at 65.8MHz, with a linewidth narrower than 10kHz.

*3.3 Stable Dual Comb Line Operation.*

A new and interesting operating regime associated with our system is presented in Fig. 5. On the timescale of the optical pulses, this operating regime is similar to that of the stable case of Fig. 4. The autocorrelation (Fig. 5 (a), red) reveals a stable pulsation consistent with the calculated autocorrelation trace (Fig. 5 (a), dashed black) obtained by assuming that the lines in the optical spectrum (Fig. 5 (b)) are in phase. However, the temporal RF signal measured with a high speed photodiode (Fig. 5(c)) along with its RF spectrum (Fig.5 (d)), shows highly coherent beating at 65MHz, indicating that two distinct modes of the main cavity are oscillating for each microring resonator line. Using a filter we verified that the beating originated from mode doublets occurring in every excited resonance of the microring.

These measurements indicate that the system was oscillating with two main cavity modes distributed around a ring resonance peak. The laser oscillated in a stable mode-locking manner with an output spectrum consisting of a number of closely spaced main cavity mode doublets. An extinction ratio of the beating (Fig. 5 (c)) that exceeds 80% demonstrates a good balance (in terms of energies) of the two comb replica. The secondary peak around 131 MHz in Fig.5 (d) is a second harmonic of the RF beat frequency, but is at a much lower power level (-23dB down with respect to the main RF peak). The inset in Fig. 5(d) shows the particularly narrow linewidth of the RF beat frequency estimated to be < 10kHz (FWHM).

This beating is related to the frequency difference between the two oscillating comb replicas and so its spectrum is an accurate reflection of the dynamics of the line-to-line spacing of the main cavity modes. This confirms the instantaneous frequency locking between the main cavity oscillating lines.

## 5. Discussion

These results demonstrate the versatility of the FD-FWM scheme as a function of the main cavity length. By adjusting the relative phase between the main cavity modes and the ring resonator, not only is it possible to eliminate all the low frequency instabilities arising from supermode beating and EDFA gain switching in order to obtain stable pulsed operation, but it is also possible to achieve stable oscillation of a dual comb separated by the FSR of the main cavity.

An important aspect of these results is that the generated RF modulation is coherent with the emission of a 200GHz optical pulse train because it arises from beating between cavity modes. This is qualitatively different from modulating the pulse train with an external modulator, for example. To first order the modulation can be thought of as a Phased Locked sub-carrier since its frequency shifts with the 200GHz rep-rate. Although it is not necessarily an integer fraction of the repetition rate, obtaining a locked sub-carrier requires one to "read" the pulse train with a Phase Locked Loop, but in this case the repetition rate (200GHz) is well beyond the capability of typical detectors. Instead, in this case the signal already carries the information that is useful, for example, to synchronize slower sources for time-division multiplexing applications. Moreover, the measurement of the amplitude noise of this configuration gives important information on the phase stability of the system, as the line width of the AC peak at 65.8MHz in Fig. 5(d) (inset) is a direct measurement of the stability of the FSR of the main cavity, which is better than 0.01% for the results presented here. This measurement reflects the stability of the ultrafast comb, and in principle can be employed to stabilize the whole system at a frequency (65.8 MHz) that can be easily followed by electronics.

## 6. Conclusions

We demonstrate a novel dual-mode locked laser based on an integrated high-Q microring resonator that exhibits stable operation of two slightly shifted spectral optical comb replicas, generating a highly monochromatic radio frequency tone.

**Acknowledgments**

This work was supported by the Natural Sciences and Engineering Research Council of Canada (NSERC) and the Australian Research Council (ARC) Discovery Projects and Centre of Excellence for Ultrahigh bandwidth Devices for Optical Systems.